\documentclass[journal]{IEEEtran}
\usepackage{amsmath,amsfonts}
\usepackage{algorithmic}
\usepackage{algorithm}
\usepackage{array}
\usepackage[caption=false,font=normalsize,labelfont=sf,textfont=sf]{subfig}
\usepackage{textcomp}
\usepackage{stfloats}
\usepackage{url}
\usepackage{verbatim}
\usepackage{graphicx}
\usepackage{cite}
\usepackage{bm}
\begin{document}

\title{Unify Variables in Neural Scaling Laws for General Audio Representations via Embedding Effective Rank}

\author{Xuyao Deng, Yanjie Sun, Yong Dou, Kele Xu,~\IEEEmembership{Senior Member,~IEEE} 
\thanks{All authors are with the College of Computer Science and Technology, National University of Defense Technology, Changsha 470000 China (e-mail: \{dengxuyao, sunyanjie21, xukelele, yongdou\}@nudt.edu.cn)}}

% The paper headers
% \markboth{Journal of \LaTeX\ Class Files,~Vol.~14, No.~8, August~2021}%
% {Shell \MakeLowercase{\textit{et al.}}: A Sample Article Using IEEEtran.cls for IEEE Journals}

% \IEEEpubid{0000--0000/00\$00.00~\copyright~2021 IEEE}
% Remember, if you use this you must call \IEEEpubidadjcol in the second
% column for its text to clear the IEEEpubid mark.

\maketitle

\begin{abstract}
Scaling laws have profoundly shaped our understanding of model performance in computer vision and natural language processing, yet their application to general audio representation learning remains underexplored. A key challenge lies in the multifactorial nature of general audio representation—representation quality is jointly influenced by variables such as audio length, embedding dimensionality, model depth, model architecture, data volume, etc., many of which are difficult to isolate or express analytically. 
In this work, we present a systematic study of scaling laws for general audio representations by utilizing \textit{embedding effective rank} (\textbf{RankMe}) as a unifying metric that encapsulates the impact of diverse variables on representation quality. RankMe enables a label-free, information-theoretic quantification of audio embeddings, allowing us to examine scaling behaviors across a wide hyper-parameter space, including model size, training data volume, computational budget, architectural configurations, etc. Our empirical findings reveal a consistent power-law relationship between RankMe and representation quality, suggesting that embedding effective rank serves as a reliable proxy for assessing and predicting model performance in audio representation learning. This work not only validates the applicability of classical scaling principles to the general audio domain but also offers a theoretically grounded and empirically robust framework for guiding future model scaling strategies in audio foundation models.
\end{abstract}


\begin{figure*}[t]
\vspace{-25pt}
    \centering
	\subfloat[]{\includegraphics[width=2.35in]{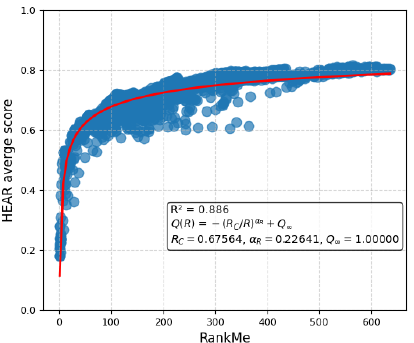}\label{fig:scaling law of RankMe-a}}
    \hfil
	\subfloat[]{\includegraphics[width=2.35in]{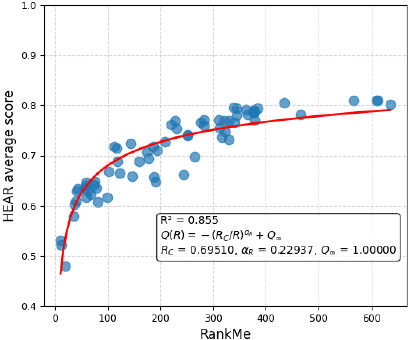}\label{fig:scaling law of RankMe with model convergence-b}}
    \hfil
	\subfloat[]{\includegraphics[width=2.35in]{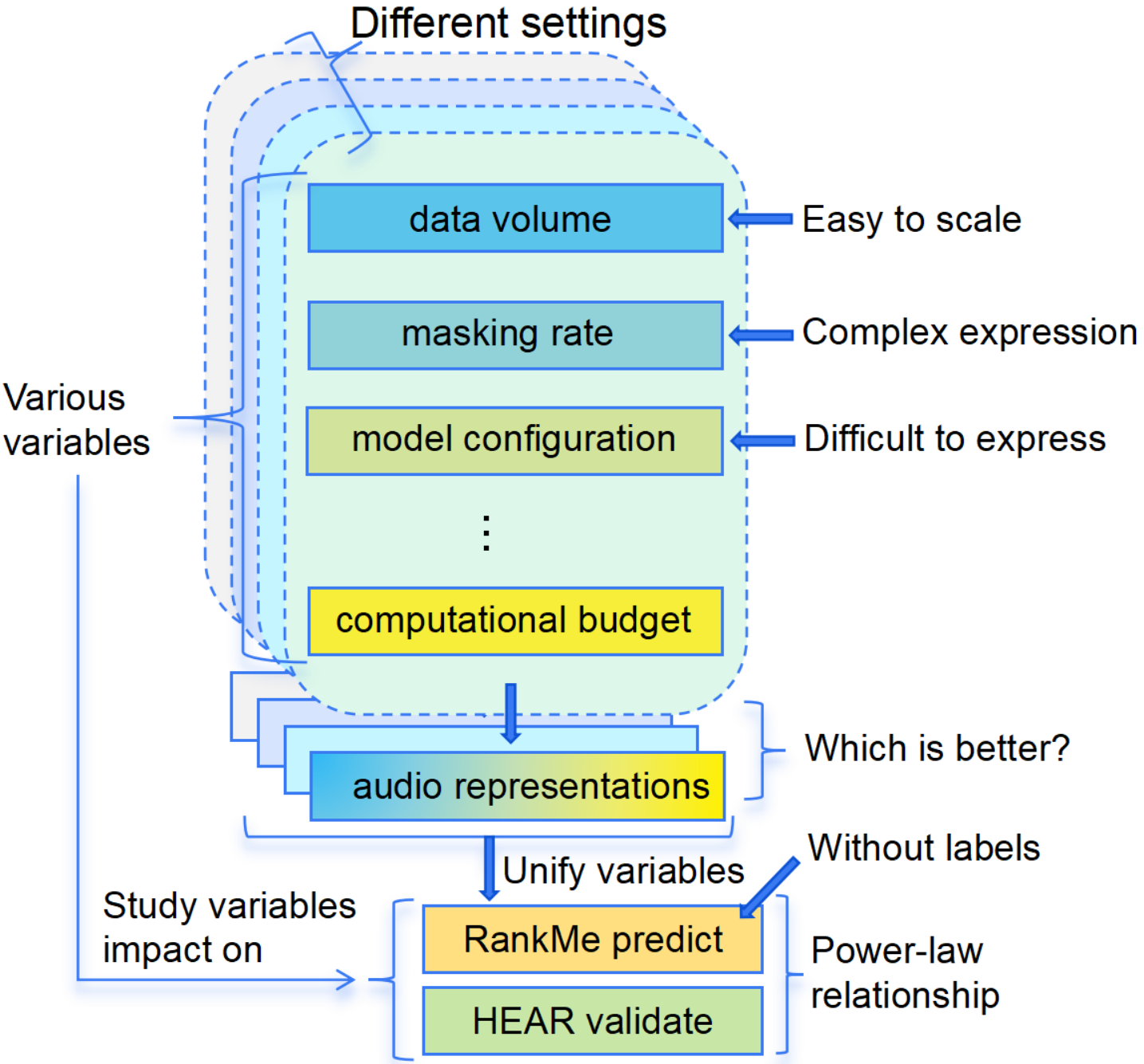}\label{fig:procedure}}  
	\caption{(a) Scaling law of embedding effective rank. (b) Convergent checkpoints. (c) Research schematic. Embedding effective rank consolidates multiple variables—including hyper-parameters that are typically difficult to express in conventional scaling laws—into a unified measurement framework. This enables the establishment of scaling laws with respect to embedding quality. Each point in (a) and (b) represents a pretrained model checkpoint under a distinct configuration, varying in data volume, model size, computational budget, masking rate, and model structure (embedding length and model depth). The key difference is that (b) includes only converged checkpoints without consideration of computational cost, while (a) reflects practical training constraints.(c) presents a schematic overview of our study. $R^2$ is coefficient of determination.}
    \label{fig:scaling law of RankMe}
\vspace{-15pt}
\end{figure*}

\section{Introduction}
\IEEEPARstart{S}{caling} laws, first proposed by~\cite{kaplan2020scaling}, have become a central research paradigm in modern machine learning. In domains such as Natural Language Processing (NLP)~\cite{isik2024scaling,hoffmann2022training,muennighoff2023scaling,aghajanyan2023scaling,isik2025scaling} and Computer Vision (CV)~\cite{fan2024scaling,klug2022scaling,cherti2023reproducible}, these laws have been extensively validated, offering insights into how model performance evolves with increased data, compute, and model capacity. In contrast, scaling laws for \textit{general audio representation learning} remain largely unexplored. While recent advances have shown the potential of self-supervised learning to extract rich audio embeddings~\cite{bengio2013representation,mohamed2022self,ericsson2022self,li2025speech}, there is still a lack of systematic empirical and theoretical investigation into how various factors jointly influence representation quality in the audio domain.

A key challenge arises from the multifactorial nature of general audio representation: representation quality is influenced by a broad range of variables, including audio duration, embedding dimensionality, model depth, masking strategy, data volume, etc. These variables often interact in complex ways and are difficult to express or scale analytically. Prior scaling law studies typically focus on a small number of independent variables that are easy to quantify. For example, Kaplan et al.~\cite{kaplan2020scaling} investigated how training loss scales individually or jointly with model size ($N$), data volume ($D$), and computational budget ($C$). However, extending this approach to general audio representations, where the effects of variables are highly entangled, presents significant modeling and analytical challenges~\cite{aghajanyan2023scaling,deng2025scaling}. In addition,~\cite{whetten2025towards} points out that the high or low loss value of pre-training cannot reflect the actual audio representation ability of the model. Moreover, variables like masking rate or model architecture exhibit irregular or non-monotonic scaling behavior, making traditional formulations insufficient~\cite{he2022masked,huang2022masked}.

Inspired by Cover's theorem~\cite{cover2006geometrical}, which states that the performance of a linear classifier improves with increasing feature rank, we posit that the \textit{effective rank} of embeddings serves as a proxy for their quality. This observation motivates us to utilize \textbf{embedding effective rank}—computed via RankMe~\cite{garrido2023rankme}—as a unified metric to characterize the collective influence of multiple variables in general audio representation learning. Unlike conventional approaches, RankMe provides an unsupervised, information-theoretic measure of representation quality that naturally accommodates both scalable and non-scalable variables. Even in the absence of labels in downstream data, RankMe can still work.

\IEEEpubidadjcol

In this work, we systematically investigate scaling laws for general audio representation through the lens of embedding effective rank. Using a masked autoencoding self-supervised learning framework~\cite{dinkel2024dasheng} mainly and other frameworks~\cite{gong2022ssast,hsu2021hubert,baevski2020wav2vec, hsu2021robust,wang2020fairseq, gulati2020conformer}, we demonstrate that effective rank scales with a power-law relationship to downstream performance, as quantified by the HEAR benchmark~\cite{turian2022hear}. Figure~\ref{fig:scaling law of RankMe}~\subref{fig:procedure} presents an overview of our methodology. Our main contributions are summarized as follows:

\begin{itemize}
    \item \textbf{Scaling Law via Embedding Effective Rank.} We show that RankMe serves as a unified measure for general audio representation quality, revealing a consistent power-law relationship under the influence of diverse factors—including those traditionally difficult to incorporate into scaling laws, such as masking rate or model architecture.

    \item \textbf{Extending RankMe to the General Audio Domain.} Originally developed for training hyper-parameters (such as learning rate) selection in JE-SSL on image tasks, we extend RankMe to general audio including various sound types such as language, ambient sound, music, etc. Our findings validate its utility in quantifying representation quality across both model-specific and general-purpose hyper-parameters, without requiring labeled data.

    \item \textbf{Empirical Analysis of Variable Impacts.} While RankMe unifies multiple variables, we investigate how individual factors—such as data volume, model size, embedding length, model depth, computational budget, and masking rate—contribute differently to effective rank. Our results uncover specific scaling trends for each.

    \item \textbf{Conventional Scaling Laws for Audio Representations.} Beyond the unified approach, we conduct a systematic study of traditional scaling laws applied to general audio representation learning, analyzing the effect of individual and joint variables on downstream performance.
\end{itemize}

\section{Related Work}

\subsection{Embedding Rank}
The conventional definition of matrix rank is often too rigid for real-world applications, particularly in deep learning, where differentiable and robust alternatives are preferable~\cite{press2007numerical}. To address the limitation that traditional rank—being a discrete integer—cannot be directly optimized (e.g., via gradient-based methods),~\cite{roy2007effective} introduced the notion of \textit{effective rank}, offering a continuous and information-theoretic approximation. Building on this concept,~\cite{garrido2023rankme} proposed \textbf{RankMe}, which incorporates Shannon entropy~\cite{shannon1948mathematical} to quantify the effective rank of an embedding matrix from a probabilistic perspective.
Originally designed for hyper-parameters selection in joint embedding self-supervised learning (JE-SSL) on image datasets~\cite{chen2020simple,bardes2021vicreg,caron2021emerging}, RankMe was also shown to correlate with model quality when varying key training parameters such as learning rate in the field of speech~\cite{aldeneh2025towards}. And~\cite{whetten2025towards} utilizes RankMe to early predict the performance of the model in the field of speech. In this work, we extend the use of RankMe to the general audio domain, leveraging its ability to compute the effective rank of general audio embeddings for analyzing scaling behavior.

\subsection{Scaling Laws}
Scaling laws describe how model performance scales with core factors such as data volume, model size, and computational budget. Foundational work by~\cite{kaplan2020scaling,henighan2020scaling,droppo21_interspeech} established the empirical relationships among these variables in NLP and speech domains. Later,~\cite{hoffmann2022training,touvron2023llama} refined this understanding by analyzing optimal trade-offs under fixed computational budgets. 
More recently,~\cite{aghajanyan2023scaling} extended scaling law analysis to multimodal settings, capturing the compound effects of multiple independent variables. However, most existing studies restrict attention to one or two easily quantifiable variables~\cite{isik2024scaling}. In contrast, our work addresses the scaling behavior under multiple interacting factors—including those difficult to express analytically—by unifying them through the lens of embedding effective rank.

\subsection{Audio Representations}
Audio representation learning aims to extract compact, semantically rich features from raw audio signals—including speech~\cite{zhu2024multichannel}, music~\cite{zhu2025muq}, and environmental sounds~\cite{shams2024ssamba}—to support downstream tasks such as classification~\cite{zaman2023survey}, segmentation~\cite{zhou2025audio,chen2019audio}, and generation~\cite{yuan2025sound,yang2023uniaudio}. A unified evaluation framework, the HEAR benchmark~\cite{turian2022hear}, has been introduced to standardize comparisons across models and tasks.
In this work, we investigate the impact of a wide range of variables—including data volume, model size, architectural factors, computational budget, and masking rate—on audio representation quality. Crucially, we demonstrate that embedding effective rank offers a unified, label-free metric to capture and analyze these interactions under a consistent scaling framework.

\section{Preliminaries}
\subsection{Pretraining Dataset}
To systematically investigate the impact of pretraining data volume on audio representation quality, we constructed a large-scale, general-purpose audio dataset comprising approximately 100 million 10-second clips (around 277,000 hours). This dataset is built by combining the unbalanced subset of AudioSet~\cite{gemmeke2017audio} with the ACAV100M corpus~\cite{lee2021acav100m}. During pretraining, we use the balanced subset of AudioSet as the validation set and monitor mean squared error (MSE) at each epoch for convergence. To simulate different data regimes, we subsample from this large corpus, scaling the training data volume from 55 hours to 277,000 hours.

\subsection{Pretraining Model}
When studying different frameworks, we used SSAST~\cite{gong2022ssast}, HuBERT~\cite{hsu2021hubert}, Wav2Vec2~\cite{baevski2020wav2vec, hsu2021robust}, and Wav2Vec2-conformer~\cite{wang2020fairseq, gulati2020conformer} architectures and followed the settings in their original papers. 

When studying hyper-parameters within a defined framework, we adopt the Dasheng architecture~\cite{dinkel2024dasheng}, which applies a masked autoencoding self-supervised learning paradigm and has achieved state-of-the-art results on the HEAR benchmark. To minimize confounding factors arising from architecture or training paradigm, we closely follow Dasheng's setup. Specifically, we use a fixed batch size of 256 and a default masking ratio of 0.75 (except when explicitly varying the mask rate from 0.3 to 0.95 in 0.05 increments). Unlike Dasheng, we process the full training set per epoch, rather than sampling. All models are trained for 746{,}000 steps unless early stopping is triggered on small-scale data to prevent overfitting. This ensures that all models, regardless of scale, observe the same total amount of training data.

\begin{figure*}[t]
\vspace{-25pt}
	\centering
	\subfloat[]{\includegraphics[width=2.35in]{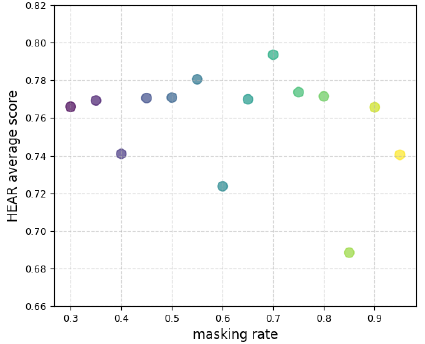}\label{fig:masking_rate-a}}
    \hfil
	\subfloat[]{\includegraphics[width=2.35in]{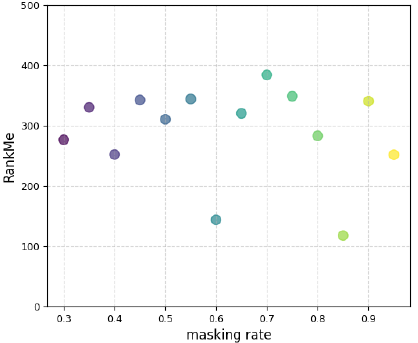}\label{fig:masking_rate-b}}
	\hfil
	\subfloat[]{\includegraphics[width=2.35in]{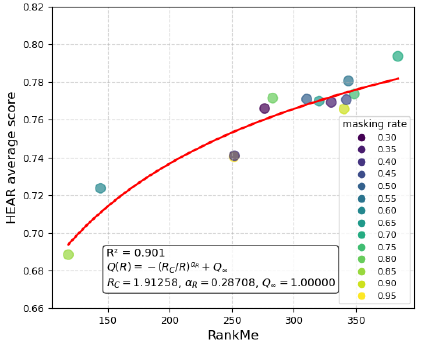}\label{fig:masking_rate-c}}
	\caption{(a) HEAR with different masking rate. (b) RankMe with different masking rate. (c) HEAR with different RankMe. The left figure shows the trend of HEAR average score with the scaling of masking rate. The middle figure shows the trend of RankMe with the scaling of masking rate. The right figure demonstrates that under different masking rate settings, RankMe and the HEAR benchmark exhibit a positive correlation conformed to Formula~\eqref{eq:Q(R)}. }
	\label{fig:masking rate}
\vspace{-15pt}
\end{figure*}

To explore the effects of compute and model scale, we vary both the training step count (from 2{,}000 to 746{,}000) and model size (from 27M to 707M parameters). The model encoder architecture is scaled by adjusting embedding length (from 128 to 1536) and depth (from 1 to 24 layers), while keeping the decoder fixed. Each encoder block uses a multi-layer perceptron (MLP) with hidden dimension four times the embedding size, and the number of attention heads is set to one sixty-fourth of the embedding length. The decoder is fixed at 512 embedding size, 8 layers, MLP dimension 2048, and 16 attention heads. Detailed architectural configurations for different model sizes are summarized in Table~\ref{tabel:Model setups}. For additional implementation details, please refer to \textbf{the Appendix} and the official Dasheng codebase~\footnote{\url{https://github.com/XiaoMi/dasheng}}.

\begin{table}[t]
\caption{Model setups. `Depth', `Embed', `MLP', `Heads' respectively represent the number of layers, the embedding length, the MLP dimension, and the number of attention heads in the encoder. All models use the same decoder.\label{tabel:Model setups}}
\centering
\begin{tabular}{|c|c|c|c|c|c|}
\hline
Model & All param & Depth & Embed & MLP & Heads \\
\hline
en1536-24 & 707.01M         & 24       & 1536     & 6144   & 24          \\
\hline
en1536-12 & 367.03M         & 12       & 1536     & 6144   & 24          \\
\hline
en1024-12 & 177.69M         & 12       & 1024     & 4096   & 16          \\
\hline
en768-12  & 111.32M         & 12       & 768      & 3072   & 12          \\
\hline
en512-12  & 63.84M          & 12       & 512      & 2048   & 8           \\
\hline
en256-12  & 35.22M          & 12       & 256      & 1024   & 4           \\
\hline
en128-12  & 27.99M          & 12       & 128      & 512    & 2           \\
\hline
en768-8   & 82.97M          & 8        & 768      & 3072   & 12          \\
\hline
en512-8   & 51.23M          & 8        & 512      & 2048   & 8           \\
\hline
en256-8   & 32.06M          & 8        & 256      & 1024   & 4           \\
\hline
en768-4   & 54.62M          & 4        & 768      & 3072   & 12          \\
\hline
en512-4   & 38.62M          & 4        & 512      & 2048   & 8           \\
\hline
en256-4   & 28.90M          & 4        & 256      & 1024   & 4           \\
\hline
en768-1   & 33.36M          & 1        & 768      & 3072   & 12          \\
\hline
en512-1   & 29.16M          & 1        & 512      & 2048   & 8           \\
\hline
en256-1   & 26.53M          & 1        & 256      & 1024   & 4           \\ 
\hline
\end{tabular}
\end{table}

\subsection{Calculating Embedding Effective Rank}
To quantify representation quality, we compute the \textit{embedding effective rank} using RankMe~\cite{garrido2023rankme}, which evaluates the rank of an embedding matrix from an information-theoretic perspective. Specifically, the singular values $\sigma_k(\bm{Z})$ of the embedding matrix $\bm{Z} \in \mathbb{R}^{D \times K}$ are normalized into a probability distribution:
{\small
\begin{equation}
    p_k = \frac{\sigma_k(\bm{Z})}{\|\sigma(\bm{Z})\|_1} + \epsilon,
\end{equation}}%
where $\epsilon$ is a small constant (typically $10^{-7}$ for \texttt{float32}). The RankMe score is then calculated as:
{\small
\begin{equation}
\label{eq:RankMe}
    \text{RankMe}(\bm{Z}) = \exp\left(-\sum_{k=1}^{\min(D,K)} p_k \log p_k \right).
\end{equation}}%
The RankMe score ranges from 1 to $K$ (the embedding length, Commonly, $K<D$), and higher values indicate richer, more uniformly distributed embeddings. Following~\cite{garrido2023rankme}, we randomly sample 30{,}000 examples from the unbalanced AudioSet training set to estimate RankMe efficiently. Importantly, the method is label-free and entirely unsupervised. In subsequent sections, we refer to embedding effective rank simply as \textbf{RankMe}.

\subsection{Downstream Tasks and Evaluation}
We evaluate the learned audio representations using the HEAR benchmark~\cite{turian2022hear}. Specifically, we freeze the upstream encoder and train a shallow MLP classifier (hidden dimension 1024) via linear probing on downstream tasks. For downstream implementation details, we refer readers to \textbf{the Appendix} and the official HEAR benchmark codebase~\footnote{\url{https://hearbenchmark.com}}.

HEAR includes two types of tasks: (1) \textbf{Scene-based} tasks (17 total), involving multi-class or multi-label classification over full audio clips, and (2) \textbf{Timestamp-based} tasks (2 total), such as event detection or transcription. In this study, we focus exclusively on Scene-based tasks. Following~\cite{anton2023audio}, we exclude the `Beehive' task due to long utterance duration and low sample count, which lead to inconsistent results. We thus evaluate on 16 Scene-based tasks: SPC-5, SPC-F, VL, LiCt, VI, CD, BJ, GZ-Gen, GZM/S, Mri-T, Mri-S, NS-5, NS-50, E50, Gun, and F50k. Final performance is reported as the average HEAR score across these 16 tasks. Task-wise performance details are provided in Section~\ref{subseq:Scaling Law of RankMe Per Task}.

\section{Results}
\subsection{Scaling Law of Embeddings Effective Rank}

To demonstrate the effectiveness of embedding effective rank in unifying variables in neural scaling laws, we obtain a diverse set of model checkpoints generated under varying hyper-parameter configurations, including data volume, model size, computational budget, masking rate, and architectural factors (e.g., embedding dimension and model depth). For each checkpoint, we compute its RankMe score and evaluate its audio representation quality using the HEAR benchmark. The results are illustrated in Figure~\ref{fig:scaling law of RankMe}, where we consider two scenarios: (a) checkpoints with non-convergent models constrained by computational cost, and (b) checkpoints with models trained to full convergence regardless of computational expense.

Our empirical findings suggest that the quality of audio representations and RankMe adhere to the following power-law relationship:
{\small
\begin{equation}
    \label{eq:Q(R)}
    Q(R) = -\left(\frac{R_C}{R}\right)^{\alpha_R}+Q_\infty,
\end{equation}
}%
where $R$ denotes the RankMe score of a model checkpoint, and $Q$ represents the quality of its audio representations, as measured by the HEAR average score. The parameters $R_C$, $\alpha_R$, and $Q_\infty$ are fitted from the data, with $Q_\infty$ representing the theoretical upper bound of representation quality, while $R_C$ and $\alpha_R$ describe the scaling behavior of $R$ required to approach that limit.

Notably, the estimated parameters differ across the two scenarios depicted in Figure~\ref{fig:scaling law of RankMe}~\subref{fig:scaling law of RankMe-a} and~\subref{fig:scaling law of RankMe with model convergence-b}. We attribute these discrepancies to the variation in the number and quality of data points used in each setting, which in turn affects the robustness of the fitted curves. Nevertheless, we hypothesize that the underlying scaling law governed by embedding effective rank reflects an invariant pattern across training configurations.

From the fitting function in Formula~\eqref{eq:Q(R)}, we derive the following key insights:
\begin{enumerate}
    \item Adjusting model hyper-parameters can improve RankMe scores, thereby enhancing the quality of audio representations. It implies that hyper-parameters adjustment can be guided by embedding effective rank especially in cases where data lacks labels and cannot be directly evaluated using data.
    \item Larger embedding dimensions may be necessary to represent audio effectively, as the RankMe score is inherently constrained when the dimensionality is too low (cf. Formula~\eqref{eq:RankMe}).
    \item Achieving peak performance would require exceedingly large RankMe scores, implying potential limitations in either data quality or the expressiveness of the masked autoencoding paradigm for self-supervised learning.
\end{enumerate}

The advantages of using embedding effective rank as a unifying axis in neural scaling laws are twofold: (1) It enables the incorporation of otherwise hard-to-formalize variables (e.g., masking rate and model architecture) into a unified scaling framework; (2) It compresses multiple heterogeneous factors into a single interpretable variable, simplifying the empirical study of scaling behaviors.

Our findings suggest that RankMe generalizes across both model-specific hyper-parameters (e.g., model size, embedding dimension, masking rate, model depth) and external factors (e.g., computational budget, data volume), positioning it as a general proxy for audio model capacity and representational power. The direct benefit it brings is that according to Formula~\eqref{eq:Q(R)}, the general audio representation ability of the model under multiple different hyperparameters can be approximately evaluated by comparing RankMe values without the need for validation on downstream tasks, which is very helpful in situations where there is no downstream task data or data without labels. In the following subsections, we further examine the effects of individual factors on RankMe and downstream performance.

\begin{figure*}[t]
\vspace{-25pt}
	\centering
    \subfloat[]{\includegraphics[width=2.35in]{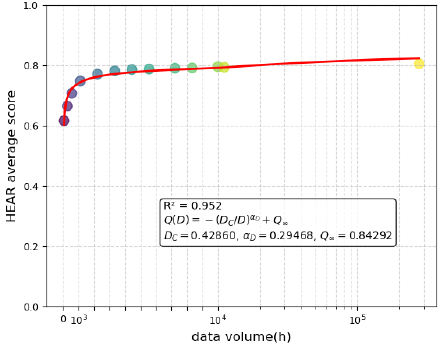}\label{fig:data_volume-a}}
    \hfil
    \subfloat[]{\includegraphics[width=2.4in]{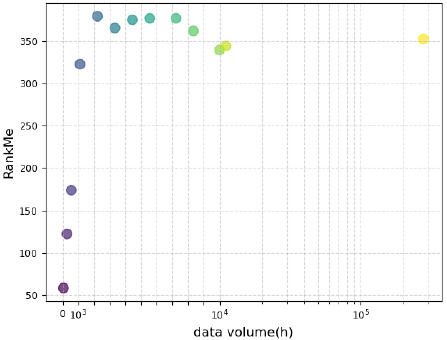}\label{fig:data_volume-b}}
    \hfil
    \subfloat[]{\includegraphics[width=2.25in]{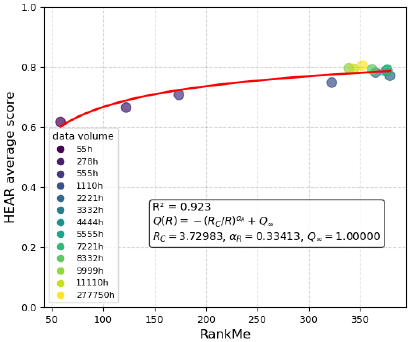}\label{fig:data_volume-c}}
    \hfil
	\caption{(a) HEAR with different data volume. (b) RankMe with different data volume. (c) HEAR with different RankMe. The left figure shows the trend of HEAR average score with the scaling of data volume. The middle figure shows the trend of RankMe with the scaling of data volume. The right figure demonstrates that under different data volume settings, RankMe and the HEAR benchmark exhibit a positive correlation conformed to Formula~\eqref{eq:Q(R)}.}
	\label{fig:data volume}
\vspace{-15pt}
\end{figure*}

\subsection{Masking Rate}
Traditional scaling laws struggle to model certain hyper-parameters for three primary reasons: (1) they are difficult to express analytically (e.g., architectural design choices or masking strategies); (2) they lack a consistent scaling dimension (e.g., loss functions); and (3) their behaviors are inherently nonlinear and complex, such as the masking rate shown in Figure~\ref{fig:masking rate}~\subref{fig:masking_rate-a}.

Figure~\ref{fig:masking rate}~\subref{fig:masking_rate-a} illustrates the relationship between masking rate and HEAR score for the en768-12 model trained on 5333 hours of audio. The trend appears non-monotonic and analytically intractable, which complicates direct integration into traditional scaling laws. However, when we express masking rate through RankMe (Figure~\ref{fig:masking rate}~\subref{fig:masking_rate-c}), a clear power-law relationship emerges, following Formula~\eqref{eq:Q(R)}. This finding demonstrates that RankMe effectively absorbs the complexity of masking rate and reveals its contribution to representation quality through a simplified scaling law.

Furthermore, Figure~\ref{fig:masking rate}~\subref{fig:masking_rate-b} aligns with the observed trends in Figure~\ref{fig:masking rate}~\subref{fig:masking_rate-a}, confirming that RankMe captures and compresses the impact of masking into a single measurable dimension. This highlights RankMe's utility in bridging complex hyper-parameter effects with performance metrics.

\subsection{Data Volume}

To explore the effect of data volume $D$ on RankMe and downstream performance, we train a series of en768-12 models under varying data sizes, with model capacity constrained to prevent saturation from other factors. We observe a consistent power-law relationship between $D$ and audio representation quality, expressed as:
{\small
\begin{equation}
    \label{eq:Q(D) = -(D_C / D)^alpha_D+Q_∞}
    Q(D) = -\left(\frac{D_C}{D}\right)^{\alpha_D}+Q_\infty,
\end{equation}
}%
where $D_C$, $\alpha_D$, and $Q_\infty$ are parameters estimated from the training results.

\begin{figure*}[t]
	\centering
    \subfloat[]{\includegraphics[width=2.35in]{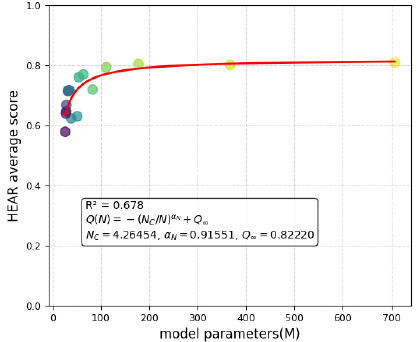}\label{fig:model_size-a}}
    \hfil
    \subfloat[]{\includegraphics[width=2.35in]{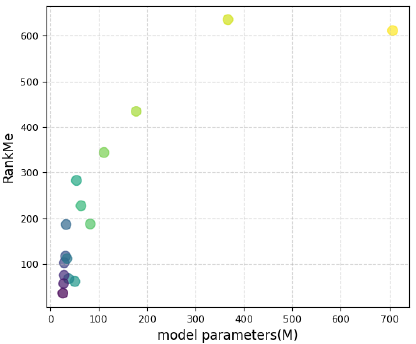}\label{fig:model_size-b}}
    \hfil
    \subfloat[]{\includegraphics[width=2.35in]{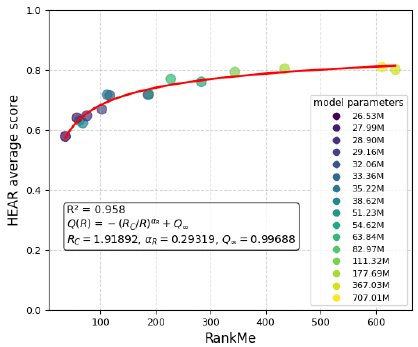}\label{fig:model_size-c}}
    \hfil
	\caption{(a) HEAR with different model size. (b) RankMe with different model size. (c) HEAR with different RankMe. The left figure shows the trend of HEAR average score with the scaling of model size. The middle figure shows the trend of RankMe with the scaling of model size. The middle figure demonstrates that under different model configuration settings, RankMe and the HEAR benchmark exhibit a positive correlation conformed to Formula~\eqref{eq:Q(R)}. }
	\label{fig:model_size}
\vspace{-15pt}
\end{figure*}

Figures~\ref{fig:data volume}~\subref{fig:data_volume-a} and~\subref{fig:data_volume-b} show how RankMe and HEAR scores grow with increasing data volume, exhibiting near-identical trends. Interestingly, in Equation~\eqref{eq:Q(D) = -(D_C / D)^alpha_D+Q_∞}, we find $Q_\infty < 1$, which we do not attribute to limitations in model size, as Figure~\ref{fig:model_size}~\subref{fig:model_size-a} shows saturation in performance for models exceeding 100M parameters trained on 10666h of data. We hypothesize that this performance ceiling is due to cumulative effects of dataset quality, pretraining strategies, and model design.

Moreover, Figure~\ref{fig:data volume}~\subref{fig:data_volume-c} demonstrates a positive correlation between RankMe and HEAR scores across different data volume, consistent with Formula~\eqref{eq:Q(R)}. This further affirms the role of RankMe as a reliable proxy for representational quality across scales. Nevertheless, we caution that RankMe remains a coarse-grained indicator—a higher value generally correlates with better performance but does not guarantee it in all cases.

\subsection{Model-specific Hyper-parameters}

To examine how model-specific hyper-parameters influence audio representation quality, we trained a series of models using 10,666 hours of audio data, as summarized in Table~\ref{tabel:Model setups}. These models were designed to be constrained primarily by architectural configurations rather than data volume. As shown in Figure~\ref{fig:model_size}~\subref{fig:model_size-a}, we observe a power-law relationship between the number of model parameters and the quality of the learned audio representations, which can be described by the following formulation:
{\small \begin{equation}
    \label{eq:Q(N) = -(N_C / N)^alpha_N+Q_∞}
    Q(N) = -\left(\frac{N_C}{N}\right)^{\alpha_N} + Q_\infty,
\end{equation}}where $N$ denotes the number of model parameters, and $N_C$, $\alpha_N$, and $Q_\infty$ are constants to be estimated, consistent with the structure of Formula~\eqref{eq:Q(R)}. Empirically, we find that $Q_\infty < 1$, indicating a saturation ceiling below perfect representation quality. Importantly, this analysis excludes data volume as a confounding factor. As shown in Figure~\ref{fig:data volume}~\subref{fig:data_volume-a}, model performance converges once the training data exceeds approximately 2,000 hours. For both model size and data volume, we observe that upon reaching a certain threshold, simply increasing the data volume or model size yields diminishing returns in both representation quality and effective rank. This observation suggests that improvements in audio representation require not just scaling, but also innovations in learning paradigms or the introduction of higher-quality training data.

Interestingly, Figure~\ref{fig:model_size}~\subref{fig:model_size-a} shows that models with similar parameter counts can yield markedly different performance. This is due to the fact that the total number of parameters serves as a coarse-grained proxy, abstracting away the specific contribution of constituent architectural choices—such as embedding dimension, network depth, MLP width, and attention heads. These internal components jointly influence the model’s representational capacity.

In contrast, Figure~\ref{fig:model_size}~\subref{fig:model_size-c} reveals that the embeddings effective rank (RankMe) exhibits a much tighter ($R^2$ is bigger) correlation with audio representation quality, following the generalized scaling law in Formula~\eqref{eq:Q(R)}. This finding underscores RankMe’s utility as a unified and sensitive metric capable of encapsulating the effects of multiple hyper-parameters in a single framework.

Additionally, Figure~\ref{fig:model_size}~\subref{fig:model_size-b} demonstrates that, when model depth is fixed, increasing the embedding dimension leads to higher RankMe values, and consequently better audio representations. According to Formula~\eqref{eq:RankMe}, while RankMe’s theoretical upper bound is constrained by the embedding size, our results show that even when comparing models with vastly different embedding dimensions—such as model \texttt{en768-1} (larger embedding) and \texttt{en128-12} (smaller embedding)—their RankMe scores and HEAR benchmark performance are consistent. This confirms that RankMe robustly captures representation quality across diverse architectural regimes, regardless of their theoretical capacity bounds.

\begin{figure*}[t]
\vspace{-25pt}
	\centering
    \subfloat[]{\includegraphics[width=2.48in]{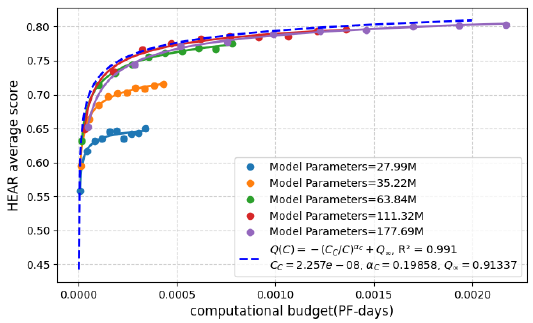}\label{fig:computational budget-a}}
    \hfil
    \subfloat[]{\includegraphics[width=2.66in]{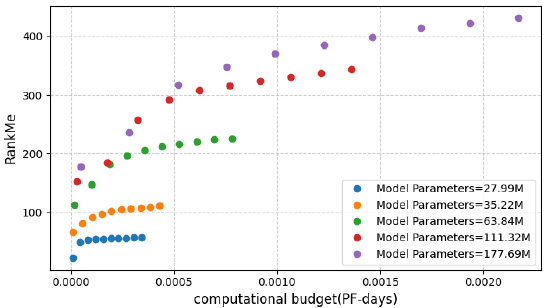}\label{fig:computational budget-b}}
    \hfil
    \subfloat[]{\includegraphics[width=1.85in]{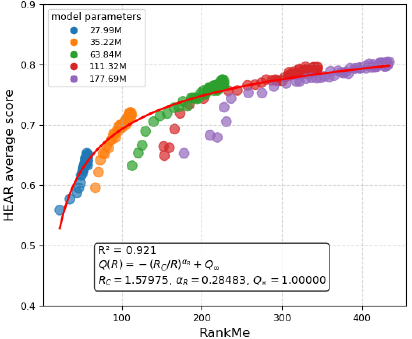}\label{fig:computational budget-c}}
    \hfil
	\caption{(a) HEAR with computational budget. (b) RankMe with computational budget. (c) HEAR with different RankMe. Scaling laws of computational budget for audio representation. The left and middle graphs respectively demonstrate the relationship between computational budget and the HEAR average score / RankMe. The right graph illustrates the positive correlation between RankMe and HEAR benchmark evaluations, which complies with Formula~\eqref{eq:Q(R)}. RankMe can quantify the quality of audio representations across varying computational budget scales.}
	\label{fig:computational budget}
\vspace{-15pt}
\end{figure*}

\subsection{Computational Budget}

To investigate the influence of computational budget on model performance, we utilized 10,666 hours of audio data. Each curve in Figure~\ref{fig:computational budget}~\subref{fig:computational budget-a} and~\subref{fig:computational budget-b} corresponds to a different model configuration characterized by a distinct number of parameters $N$. The horizontal axis of each data point represents the approximate number of multiply-add operations performed during training, following the conventions outlined in~\cite{kaplan2020scaling}.

In Figure~\ref{fig:computational budget}~\subref{fig:computational budget-a}, we observe that, irrespective of model size, none of the experimental results surpass the dashed line representing a power-law plus constant fit. This curve denotes the computational efficiency frontier and is described by the following equation, which is structurally analogous to Formula~\eqref{eq:Q(D) = -(D_C / D)^alpha_D+Q_∞} and~\eqref{eq:Q(N) = -(N_C / N)^alpha_N+Q_∞}:
{\small
\begin{equation}
    \label{eq:Q(C) = -(C_C / C)^alpha_C+Q_∞}
    Q(C) = -\left(\frac{C_C}{C}\right)^{\alpha_C} + Q_\infty,
\end{equation}}%
where $C_C$, $\alpha_C$, and $Q_\infty$ are hyper-parameters estimated via regression. Notably, $Q_\infty < 1$, suggesting that even with unlimited computational resources, the upper bound of audio representation quality remains inherently constrained.

In both Figure~\ref{fig:computational budget}~\subref{fig:computational budget-a} and~\subref{fig:computational budget-b}, a consistent trend emerges: models with larger parameter counts achieve higher HEAR average scores and RankMe values with fewer computational operations, whereas smaller models require significantly more computational effort to reach comparable performance levels. This finding suggests that training larger models for fewer steps is a more efficient strategy for increasing the effective rank of embedding and improving the quality of audio representations.

%如图5-c所示，随着训练步长的增加，各模型的RankMe值均呈现上升趋势，HEAR指标同样逐步增长。值得关注的是，同一训练步长下，不同模型的RankMe值存在显著差异。表1表明，在训练初期相同步长下RankMe值较高的模型，在后续训练过程中往往展现出更优的表征能力。这暗示着可利用模型预训练早期RankMe值预测其后期表征能力。为了验证这一观点，我们系统开展了实验：通过调整掩码率、数据量、模型参数量、模型宽度及长度等多组超参数，构建了45种不同的模型组合，并计算了这些模型在训练早期50k、100k、200k及300k步长时的RankMe值，与训练后期700k步长时HEAR指标的皮尔逊相关系数（见表2）。表2结果显示，模型早期步长的RankMe值与训练后期表征能力（以HEAR指标衡量）之间存在较强的正相关；并且随着所取早期step的延后（即计算RankMe时的训练进度推进），这种相关性呈现逐步增强的趋势。这表明，利用训练前期计算的RankMe值对模型后期表征能力进行预筛选，能够有效缓解大规模预训练中的计算资源压力，避免对所有模型均完成完整预训练流程。

As shown in Figure~\ref{fig:computational budget}~\subref{fig:computational budget-c}, with the increase of training steps, the RankMe values of each model show an upward trend, and the HEAR values also gradually increase. It is worth noting that there are differences in RankMe values among different models under the same training step. Table~\ref{tabel:RankMe and HEAR values of different models at different steps} shows that models with higher RankMe values at the same step size in the early stages of training often exhibit better audio representation abilities in subsequent training processes. This suggests that the RankMe values from the early stages of model pre-training can be used to predict its later audio representation ability. To verify this viewpoint, we conducted experiments systematically: by adjusting multiple sets of hyperparameters such as mask rate, data volume, model parameter volume, model width, and length, we constructed 45 models with different combinations of hyperparameters, and calculated the RankMe values of these models at 50k, 100k, 200k, and 300k steps in the early stages of training, as well as the Pearson correlation coefficient with the HEAR value at 700k steps in the later stages of training (see Table~\ref{tabel:Pearson correlation coefficient}). Table~\ref{tabel:Pearson correlation coefficient} shows that there is a strong positive correlation between the RankMe value in the early stages of the model and the audio representation ability (measured by the HEAR metric) in the later stages of training; and as the number of steps in calculating RankMe increases, this correlation shows a gradually increasing trend. This indicates that using the RankMe value calculated in the early stage of training to pre-screen the audio representation ability of the model in the future can effectively alleviate the computational resource pressure in large-scale pre-training and avoid completing the complete pre-training process for all models.

\begin{table}[t]
\caption{RankMe and HEAR values of different models at different steps.\label{tabel:RankMe and HEAR values of different models at different steps}}
\centering
\begin{tabular}{|c|c|c|}
\hline
Model & step100k(RankMe/HEAR) & step700k(RankMe/HEAR)\\
\hline
en768-12 & 201.865 / {\scriptsize 0.744}   &  {\scriptsize 342.007} / 0.793\\
\hline
en1024-12 & \textbf{258.111} / {\scriptsize 0.753}   &  {\scriptsize 432.298} / \textbf{0.800}\\
\hline
en512-12 & 156.360 / {\scriptsize 0.719}   &  {\scriptsize 226.555} / 0.765\\
\hline
en256-12 & 83.371 / {\scriptsize 0.662}   &  {\scriptsize 111.505} / 0.720\\
\hline
en128-12 & 50.200 / {\scriptsize 0.619}   &  {\scriptsize 56.892} / 0.633\\
\hline
\end{tabular}
\end{table}

\begin{table}[t]
\caption{Pearson correlation coefficient (PCC) between RankMe and HEAR at different steps.\label{tabel:Pearson correlation coefficient}}
\centering
\begin{tabular}{|c|c|c|}
\hline
step4RankMe & step4HEAR & PCC\\
\hline
50k & 700k &  0.636 \\
\hline
100k & 700k &  0.790 \\
\hline
200k & 700k &  0.855 \\
\hline
300k & 700k &  0.862 \\
\hline
\end{tabular}
\end{table}

Furthermore, Figure~\ref{fig:computational budget}~\subref{fig:computational budget-c} demonstrates that the relationship between RankMe and audio representation quality closely follows Formula~\eqref{eq:Q(R)}. This further validates the utility of RankMe as a robust metric for quantifying audio representation quality across varying computational budgets.

\begin{figure*}[t]
\vspace{-25pt}
	\centering
    \subfloat[]{\includegraphics[width=2.55in]{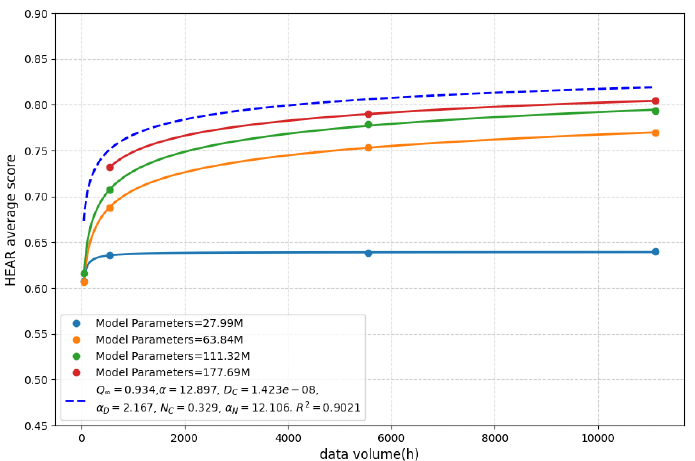}\label{fig:Model parameters and data volume-a}}
    \hfil
    \subfloat[]{\includegraphics[width=2.4in]{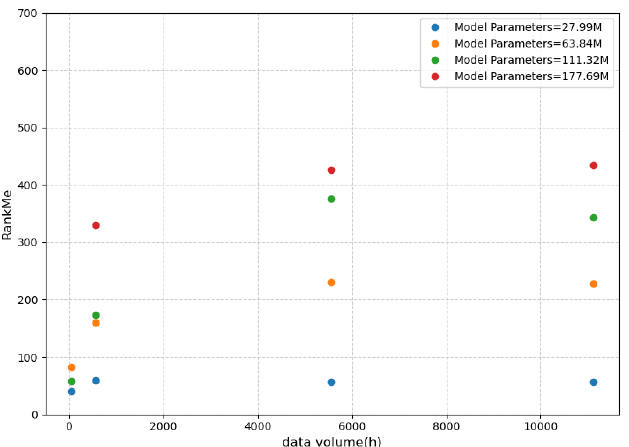}\label{fig:Model parameters and data volume-b}}
    \hfil
    \subfloat[]{\includegraphics[width=2.1in]{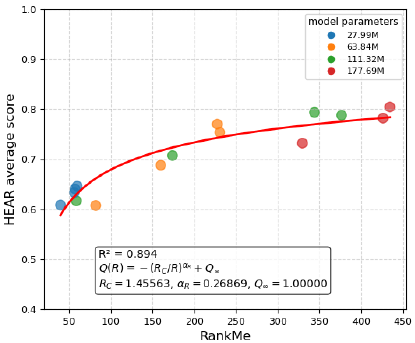}\label{fig:Model parameters and data volume-c}}
    \hfil
	\caption{(a) HEAR with data volume and model size. (b) RankMe with data volum and model size. (c) HEAR with different RankMe. Scaling laws of data volume and model size for audio representation. The left graphs shows the scaling laws of audio representations constrained by both data volume and model size, which complies with Formula~\eqref{eq:Q(N,D)}.The middle graph displays the combined effect of data volume and model size on RankMe. The right graph illustrates RankMe can simultaneously quantify the quality of audio representations across different scales of data volume and model size.}
	\label{fig:Model parameters and data volume}
\vspace{-15pt}
\end{figure*}

\subsection{Model Size and Data Volume}

In~\cite{kaplan2020scaling}, the authors demonstrated that when computational resources are unconstrained, the scaling laws with respect to model size and data volume—namely, Formula~\eqref{eq:Q(D) = -(D_C / D)^alpha_D+Q_∞} and~\eqref{eq:Q(N) = -(N_C / N)^alpha_N+Q_∞}—can be unified into a single joint scaling formulation. To incorporate the concept of an upper bound on representation quality, $Q_\infty$, we adopt the generalized formulation proposed in~\cite{droppo21_interspeech}, leading to the following composite scaling law:
{\small
\begin{equation}
    \label{eq:Q(N,D)}
    Q(N, D) = \left[\left(Q_\infty\right)^{\frac{1}{\alpha}} + \left(\frac{N_C}{N} \right)^{\frac{\alpha_N}{\alpha}} + \left(\frac{D_C}{D} \right)^{\frac{\alpha_D}{\alpha}}\right]^{\alpha},
\end{equation}}%
where $Q_\infty$, $\alpha$, $N_C$, $\alpha_N$, $D_C$, and $\alpha_D$ are hyper-parameters that govern the scaling behavior of model parameters $N$ and data volume $D$ in determining audio representation quality.

Figure~\ref{fig:Model parameters and data volume}~\subref{fig:Model parameters and data volume-a} illustrates the empirical scaling trends when both data volume and model size are jointly constrained. The dashed line indicates the performance bottleneck caused by training data volume limitations, which model performance cannot improve regardless of increased model capacity. This suggests the existence of a data-volume-induced upper bound on audio representation quality.

Compared to Formula~\eqref{eq:Q(N) = -(N_C / N)^alpha_N+Q_∞} and~\eqref{eq:Q(D) = -(D_C / D)^alpha_D+Q_∞}, the joint formulation in Formula~\eqref{eq:Q(N,D)} is more complex due to the introduction of multiple interacting variables. As the number of such variables increases, the mathematical formulation may become increasingly intricate and less interpretable. 

To address this complexity, we analyze how the RankMe metric encapsulates the effects of both model and data scale within a unified framework. Figure~\ref{fig:Model parameters and data volume}~\subref{fig:Model parameters and data volume-c} shows that RankMe scaling under joint constraints of $N$ and $D$ still conforms to the simpler power-law form of Formula~\eqref{eq:Q(R)}, suggesting its robustness and generalizability. Furthermore, Figure~\ref{fig:Model parameters and data volume}~\subref{fig:Model parameters and data volume-b} displays the combined influence of model size and data volume on RankMe, exhibiting trends consistent with those observed in Figure~\ref{fig:Model parameters and data volume}~\subref{fig:Model parameters and data volume-a}. Together, Figures~\ref{fig:Model parameters and data volume}~\subref{fig:Model parameters and data volume-b} and~\subref{fig:Model parameters and data volume-c} demonstrate that RankMe reliably quantifies audio representation quality in the presence of multiple interacting factors, including both model scale and data availability.

\subsection{Pre-training Architectures}

In addition to studying the Dasheng pre-training architecture, we also explored the scaling laws of different architectures on Rankme. We have chosen the SSAST~\cite{gong2022ssast}, HuBERT~\cite{hsu2021hubert}, Wav2Vec2~\cite{baevski2020wav2vec, hsu2021robust}, and Wav2Vec2-conformer~\cite{wang2020fairseq, gulati2020conformer} architectures and set different parameters. As shown in Table~\ref{tabel:architectures pre-training settings}, different architectures have different hyperparameter settings, including different pre-training methods, different loss functions, different masking methods, different pre-training datasets, and different model sizes. Please refer to the respective architecture papers for the parameter settings corresponding to the names of each model.

\begin{table}[t]
\vspace{-15pt}
\caption{Pre-training architectures settings. The symbol * indicates the use of pre-trained weights publicly available in the papers. LL, AS, CV, SB, F, LS, and LV represent the LibriLight, AudioSet, CommonVoice, Switchboard, Fisher, LibriSpeech, and LibriVox datasets, respectively. ACAV2M represents a 2M dataset randomly sampled from ACAV100M.\label{tabel:architectures pre-training settings}}
\centering
\begin{tabular}{|c|c|c|}
\hline
Architecture & Model & Pre-training dataset \\
\hline
SSAST\textsuperscript{*} & Tiny-Frame-400 (6M)      & AS2M+LS \\
\hline
SSAST\textsuperscript{*} & Tiny-Patch-400 (6M)      & AS2M+LS \\
\hline
SSAST\textsuperscript{*} & Small-Patch-400 (23M)      & AS2M+LS \\
\hline
SSAST\textsuperscript{*} & Base-Frame-250 (89M)      & AS2M+LS \\
\hline
SSAST\textsuperscript{*} & Base-Frame-400 (89M)      & AS2M+LS \\
\hline
SSAST\textsuperscript{*} & Base-Patch-250 (89M)      & AS2M+LS \\
\hline
SSAST\textsuperscript{*} & Base-Patch-400 (89M)      & AS2M+LS \\
\hline
Wav2Vec2-conformer\textsuperscript{*} & Large-rel-pos (317M)      & LS \\
\hline
Wav2Vec2-conformer\textsuperscript{*} & Large-rope (317M)      & LS \\
\hline
HuBERT & Base (95M)      & AS2M+ACAV2M \\
\hline
HuBERT & Large (317M)      & AS2M+ACAV2M \\
\hline
HuBERT & X-large (964M)      & AS2M+ACAV2M \\
\hline
Wav2Vec2 & Base (95M)      & AS2M+ACAV2M \\
\hline
Wav2Vec2 & Large (317M)      & AS2M+ACAV2M \\
\hline
Wav2Vec2\textsuperscript{*} & Large-lv60 (317M)      &  LV\\
\hline
Wav2Vec2\textsuperscript{*} & Large-robust (317M)      &  LL+CV+SB+F\\
\hline
Dasheng & en768-12 (111M Base)      &  AS2M+ACAV2M\\
\hline
Dasheng & en1024-12 (177M)      &  AS2M+ACAV2M\\
\hline
Dasheng & en1536-12 (367M Large)      &  AS2M+ACAV2M\\
\hline
Dasheng & en1536-24 (707M)      &  AS2M+ACAV2M\\
\hline
Dasheng & en1024-12 (177M)      &  AS2M\\
\hline
Dasheng & en1536-12 (367M)      &  AS0.2M\\
\hline
\end{tabular}
\end{table}

\begin{figure}[t] % 位置说明符：h(当前位置)、t(顶部)、b(底部)、p(单独一页)
  \centering % 图片居中
  \includegraphics[width=0.45\textwidth]{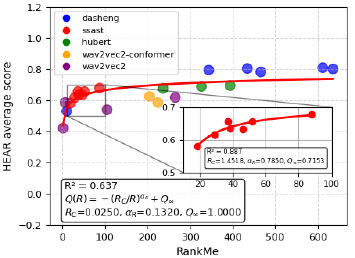} % 插入图片（宽度为文本的60%）
  \caption{HEAR with different RankMe under different architectures and settings. The locally enlarged part is separately fitted with data points under the SSAST architecture.} % 标题（自动生成编号）
  \label{fig:diffier_models} % 标签（用于交叉引用）
\vspace{-15pt} 
\end{figure}

\begin{table}[t]
\vspace{-15pt}
\caption{RankMe and HEAR values for different architecture models at different similar scales.\label{tabel:different architecture scale}}
\centering
\begin{tabular}{|c|c|c|}
\hline
Architecture & Base (RankMe / HEAR) & Large (RankMe / HEAR)\\
\hline
Wav2Vec2 & 105.153 / {\scriptsize 0.539}   &  {\scriptsize 265.285} / 0.618\\
\hline
HuBERT & 237.024 / {\scriptsize 0.677}      & {\scriptsize 326.596} / 0.689 \\
\hline
Dasheng & \textbf{344.012} / {\scriptsize 0.794}     & {\scriptsize 635.651} / \textbf{0.801} \\
\hline
\end{tabular}
\end{table}

Figure~\ref{fig:diffier_models} shows the variation of audio representation capability with RankMe under different architectures and settings. As shown in Figure~\ref{fig:diffier_models}, even across different architectures and various parameter settings, RankMe still exhibits a power-law pattern in the evaluation of general audio representation ability. In addition, from Figure~\ref{fig:diffier_models}, we found that under the same architecture but different parameter settings, each framework exhibits its own scaling trend regarding RankMe. This suggests that, as shown in Table~\ref{tabel:different architecture scale}, when we need to choose models with different architectures for scaling up and the downstream data of the application is unlabeled, we can first compare the RankMe metrics of different architectures on smaller scale models and prioritize selecting architectures with higher RankMe values for extension. This strategy enables us to screen out architectures with greater scalability potential even without the need for direct evaluation of downstream data, effectively avoiding the high computational overhead of expanding and pre-training each candidate architecture.

\subsection{Scaling Law of RankMe Per Task}
\label{subseq:Scaling Law of RankMe Per Task}
\begin{figure*}[t] % 位置说明符：h(当前位置)、t(顶部)、b(底部)、p(单独一页)
\vspace{-15pt}
  \centering % 图片居中
  \includegraphics[width=\textwidth]{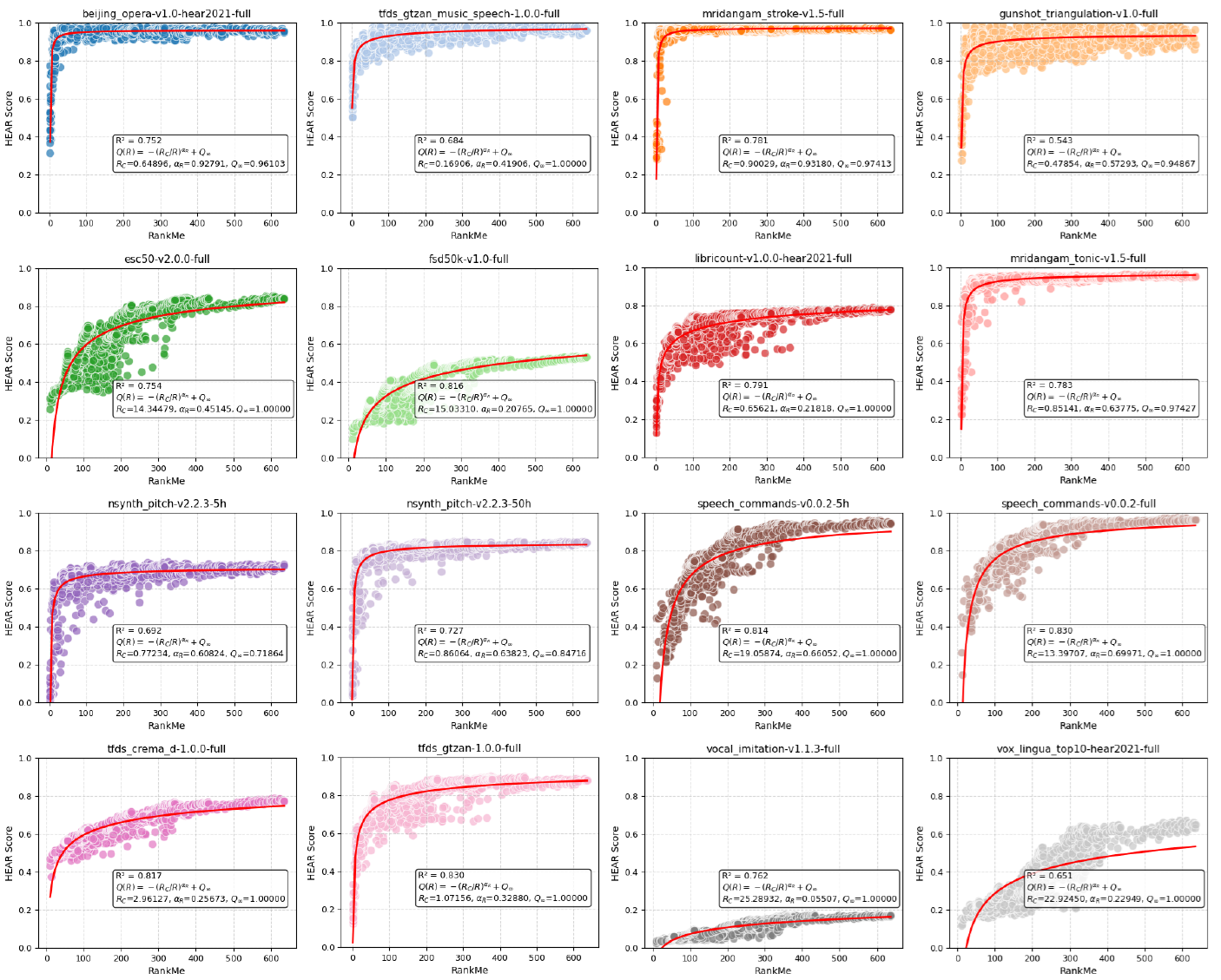} % 插入图片（宽度为文本的60%）
  \caption{Scaling law of RankMe for each task.} % 标题（自动生成编号）
  \label{fig:all_tasks_RankMe} % 标签（用于交叉引用）
\vspace{-15pt}
\end{figure*}

During the experiment, we used the average performance of 16 downstream classification tasks in the HEAR benchmark. In order to avoid masking the scaling pattern of the performance of each task with changes in RankMe, we conducted individual analyses for each task. As illustrated in Figure~\ref{fig:all_tasks_RankMe}, we present the impact of scaled RankMe on 16 downstream tasks within the HEAR benchmark. 

For different downstream tasks, the hyper-parameters in Formula~\eqref{eq:Q(R)} estimated vary, which stems from the inherent differences in the properties of each task. The value of $Q_\infty$ reflects the performance limit of the task. Parameters $R_C$ and $\alpha_R$ characterize the influence of scaling RankMe on task performance: specifically, a larger $R_C$ and a smaller $\alpha_R$ together indicate weaker sensitivity to the scaling of RankMe. $Q_\infty$, $R_C$, and $\alpha_R$ together indicate the difficulty level of the task. The smaller $Q_\infty$ is, the larger $R_C$ is, and the smaller $\alpha_R$ is, the more difficult the task becomes. We divide the 16 downstream tasks into two types: tasks that can easily achieve good performance with a lower RankMe value, but as the RankMe value increases, the performance almost reaches the horizontal line, including beijing\_opera-v1.0-hear2021-full, tfds\_gtzan\_music\_speech-1.0.0-full, mridangam\_stroke-v1.5-full, gunshot\_triangulation-v1.0-full, mridangam\_tonic-v1.5-full, nsynth\_pitch-v2.2.3-5h and nsynth\_pitch-v2.2.3-5h. And tasks that are not very easy to achieve good performance, but as the RankMe value increases, the performance gradually improves and $Q_{\infty}=1$. We believe that the reason why the 16 different tasks exhibit different types is that the first type of task usually has fewer categories, simpler audio, and is easier for the model to learn and reach performance bottlenecks, while the second type of task is the opposite; it usually has more complex audio and more categories, and the model needs to work hard to learn in order to achieve good performance.

Through experiments, we found that various downstream tasks, including speech, environmental sounds, music, language, animal sounds, etc., all exhibit power-law patterns with RankMe. RankMe exhibits stable power-law patterns across multiple different datasets.

% \begin{figure}[t] % 位置说明符：h(当前位置)、t(顶部)、b(底部)、p(单独一页)
%   \centering % 图片居中
%   \includegraphics[width=0.48\textwidth]{Figure/all_tasks_RankMe_additive_cut.pdf} % 插入图片（宽度为文本的60%）
%   \caption{Scaling law of RankMe for six tasks under harsh conditions} % 标题（自动生成编号）
%   \label{fig:all_tasks_RankMe_additive_cut} % 标签（用于交叉引用）
% \end{figure}

\section{Conclusion}

In this study, we demonstrated that embeddings effective rank serves as a unifying metric that consolidates diverse variables into a fixed, consistent perspective for analyzing scaling laws in general audio representation learning. By leveraging embeddings effective rank, we successfully incorporated hyper-parameters that are traditionally challenging to include in scaling law analyses, enabling the exploration of scaling behaviors under the joint influence of multiple factors.
Our empirical results reveal that the impact of various variables on embeddings effective rank aligns closely with patterns observed in classical scaling laws. Given that embeddings effective rank effectively captures the combined effects of different factors on audio representation quality, our findings suggest its potential as a principled guiding metric for designing and optimizing audio representation learning methods—beyond the conventional emphasis on simply scaling model size or training data volume.
Future work will explore the generalization of embeddings effective rank as a unifying metric across diverse pretraining frameworks and modalities, aiming to further deepen the theoretical understanding and practical utility of scaling laws in representation learning.

\appendices
\section{Sampling in RankMe Calculation}
\begin{figure}[htbp] % 位置说明符：h(当前位置)、t(顶部)、b(底部)、p(单独一页)
\vspace{-15pt}
  \centering % 图片居中
  \includegraphics[width=0.36\textwidth]{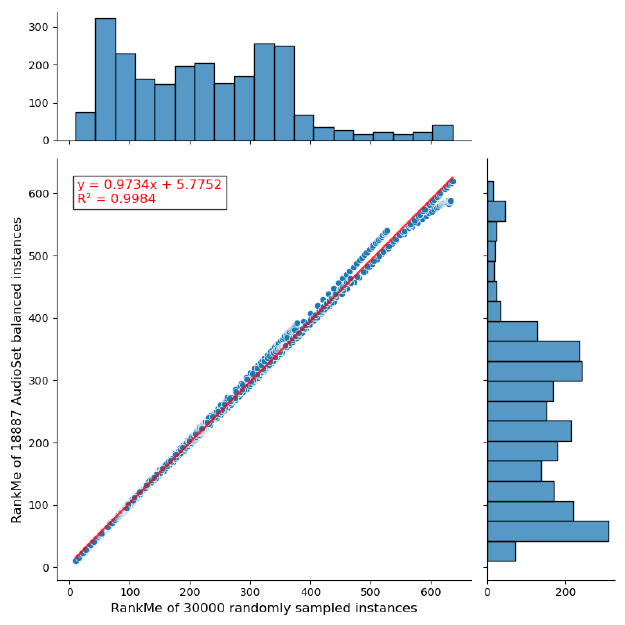} % 插入图片（宽度为文本的60%）
  \caption{Sampling in RankMe calculation.} % 标题（自动生成编号）
  \label{fig:RankMe_statistics} % 标签（用于交叉引用）
\vspace{-15pt}
\end{figure}

In our experiments, we compute RankMe using 30,000 randomly sampled data points from the training dataset. To investigate the impact of sampling, we use the full set of 18,887 data points from the AudioSet balanced dataset as an anchor and compare RankMe values calculated from two sources: the 30,000 randomly sampled data points at each checkpoint, and the 18,887 data points from AudioSet balanced dataset. As shown in Figure~\ref{fig:RankMe_statistics}, we observe negligible differences between the RankMe values derived from the 18,887 AudioSet balanced data points and those from the 30,000 randomly sampled data points, which corroborates the findings reported in (Garrido et al. 2023). Additionally, we find that RankMe values computed using 18,887 data points and those using 30,000 data points exhibit negligible variation at the same order of magnitude, despite differing in numbers.

\section{Details of Pre-training}

We followed Dasheng's setup. Specifically, our pre-training model adopts a masked autoencoder structure, incorporates learnable absolute positional embeddings, and produces frame-level embeddings at a higher frequency of 25 Hz. It operates on consecutive chunks of Mel-spectrogram frames.

In terms of data processing, we resampled all datasets to 16 kHz and extracted 64-dimensional log-Mel spectrograms at 10ms intervals with a window size of 32ms, processing 10-second audio clips. During pre-training, a grouped masking strategy was implemented to prevent the last frame of the Mel spectrogram from containing future frame information. To remain consistent with Dasheng and MAE, we set the masking ratio to 0.75. Model training employs an 8-bit AdamW optimizer combined with a cosine decay scheduler. The initial learning rate is 0.0003, and the weight decay rate is 0.01. We incorporate a learning rate warmup phase of 3 epochs, after which the learning rate is decayed to 10\% of its maximum value during the training period. The neural network back-end is implemented in PyTorch.

In the experimental process, we consistently set the batch size to 256. Different from Dasheng, we do not sample from the dataset during one epoch of pre-training. Instead, each epoch processes the complete pre-training dataset. For all model sizes and dataset scales during pre-training, unless early stopping was triggered to prevent overfitting on small datasets, the number of steps was uniformly set to 746,000. This ensures that across all experiments, models ultimately processed the same total amount of data (including multiple passes over the same data instance) by the end of pre-training, and that pre-training had converged.

During our experiments, we controlled computational budget by incrementally scaling both the step size and the model size. Specifically, the step size was scaled from 2,000 to 746,000, and the model size was scaled from 27M to 707M parameters. When scaling the model size, we kept the decoder architecture fixed and controlled the total parameter count by adjusting the width (embedding length) and depth (number of layers) of the encoder. We scaled the encoder's embedding length from 128 to 1536 and its depth from 1 to 24 layers. Throughout these changes, the blocks' multilayer perceptron dimension within encoder module was maintained at four times the embedding length, and the number of attention heads in the encoder was set to one sixty-fourth (1/64) of the embedding length. The decoder maintained a fixed embedding length of 512, a depth of 8 layers, an MLP dimension of 2048 within each module, and 16 attention heads. For specific model implementation details and the pre-training procedure, we refer the reader to the source code made public by the Dasheng authors~\footnote{\url{https://github.com/XiaoMi/dasheng}}.

\section{Details of Downstream Evaluation}
A good representation should (1) transfer to a wide range of different tasks and (2) transfer with limited supervision. Therefore, when evaluating on downstream tasks, we follow the setup of the HEAR benchmark to quantify the quality of the model's audio representations. Specifically, we freeze the parameters of the upstream pre-trained model and perform downstream task evaluation using linear probing only. The downstream model is a multilayer perceptron. Each hidden layer in MLP has a dimension of 1024, followed by BatchNorm for normalization, a Dropout layer with a rate of 0.1, and finally activated by a ReLU function.

For each dataset, during downstream task evaluation, we perform a grid search over: learning rates in [3.2e-3, 1e-3, 3.2e-4, 1e-4], number of MLP layers in [1, 2] and initialization methods of Xavier initialization following a uniform or normal distribution. The parameter configuration achieving the best performance on the validation set is selected and subsequently evaluated on the test set. During training, we employ early stopping with a maximum epoch setting of 500. Training halts when the validation loss fails to decrease for 20 consecutive epochs. For multi-label tasks, the final layer of the MLP uses the Sigmoid activation function and the BCE (binary cross entropy) loss. For multi-class tasks, the final layer of the MLP uses the Softmax activation function and the CE (cross entropy) loss. For more detailed content regarding downstream model, we refer readers to the public code implementation of the HEAR benchmark~\footnote{~\url{https://hearbenchmark.com}}

\section{Details of dataset}
We use AudioSet and ACAV100M as the pretraining datasets for our masked autoencoding self supervised learning, and 16 tasks (SPC-5, SPC-F, VL, LiCt, VI, CD, BJ, GZ-Gen, GZM/S, Mri-T, Mri-S, NS-5, NS-50, E50, Gun, and F50k.) from HEAR benchmark as the datasets for evaluating audio representation quality downstream.

\subsection{AudioSet}
AudioSet is a landmark, large-scale audio dataset released by Google in 2017 to advance research in audio event recognition and sound understanding. Designed to address the growing need for robust audio analysis in machine learning, it has since become a cornerstone resource for training and evaluating models in tasks such as sound classification, detection, and segmentation.

AudioSet consists of an expanding ontology of 632 audio event classes and a collection of 2,084,320 human-labeled 10-second sound clips drawn from YouTube videos. The ontology is specified as a hierarchical graph of event categories, covering a wide range of human and animal sounds, musical instruments and genres, and common everyday environmental sounds. We recommend readers to visit AudioSet's official homepage~\footnote{\url{https://research.google.com/audioset/index.html}}.

\subsection{ACAV100M}
ACAV100M is a landmark, fully automated dataset engineered for advancing audio-visual representation learning, unveiled by Lee et al. at the 2021 International Conference on Computer Vision (ICCV). This dataset represents a quantum leap in scale and methodology, forged by processing an astounding 140 million full-length videos—equivalent to 1,030 years of content—through a novel optimization pipeline. By prioritizing clips that maximize mutual information between audio and visual channels, the team curated 100 million 10-second segments (spanning ~31 years), dwarfing prior benchmarks like AudioSet (8 months) and HowTo100M (15 years). Key to its success is an unsupervised clustering approach (using 500 centroids) that ensures contextual relevance while filtering noise at industrial efficiency. We recommend readers to visit ACAV100M's official homepage~\footnote{\url{https://acav100m.github.io/}}.

\subsection{Speech Commands v2, 5h and full (SPC-5, SPC-F)}
Speech Commands is an audio dataset of spoken words designed to help train and evaluate keyword spotting systems. Its primary goal is to provide a way to build and test small models that detect when a single word is spoken, from a set of ten target words, with as few false positives as possible from background noise or unrelated speech. Each audio file contains a single spoken English word or background noise. These words are from a small set of commands, and are spoken by a variety of different speakers. We recommend readers to read the paper presenting Speech Commands~\footnote{\url{https://arxiv.org/abs/1804.03209}}.

\subsection{NSynth Pitch (NS-5, NS-50)}
The NSynth Dataset is an audio dataset containing about 300k musical notes, each with a unique pitch, timbre, and envelope. Each note is annotated with three additional pieces of information based on a combination of human evaluation and heuristic algorithms: Source, Family, and Qualities. In HEAR benchmark, NSynth Pitch tasks means pitch classification of synthesized sounds. We recommend readers to visit NSynth's official homepage~\footnote{\url{https://magenta.tensorflow.org/datasets/nsynth}}.

\subsection{Beijing Opera Percussion (BJ)}
The Beijing Opera percussion instrument dataset is a collection of audio examples of individual strokes spanning the four percussion instrument classes used in Beijing Opera. Beijing Opera uses six main percussion instruments that can be grouped into four classes: Bangu (Clapper-drum) consisting of Ban (the clapper, a wooden board-shaped instrument) and danpigu (a wooden drum struck by two wooden sticks); Naobo (Cymbals) consisting of two cymbal instruments Qibo and Danao; Daluo (Large gong) and Xiaoluo (Small gong). We recommend readers to visit Beijing Opera Percussion's official homepage~\footnote{\url{https://compmusic.upf.edu/bo-perc-dataset}}.

\subsection{CREMA-D (CD)}
CREMA-D is an audio-visual data set for emotion recognition. The dataset consists of facial and vocal emotional expressions in sentences spoken in a range of basic emotional states (happy, sad, anger, fear, disgust, and neutral). 7,442 clips of 91 actors (from 48 male and 43 female actors between the ages of 20 and 74) with diverse ethnic backgrounds (African America, Asian, Caucasian, Hispanic, and Unspecified) were collected. We recommend readers to visit CREMA-D's official homepage~\footnote{\url{https://github.com/CheyneyComputerScience/CREMA-D}}.

\subsection{ESC-50 (E50)}
The ESC-50 dataset is a labeled collection of 2000 environmental audio recordings suitable for benchmarking methods of environmental sound classification. The dataset consists of 5-second-long recordings organized into 50 semantical classes (with 40 examples per class) loosely arranged into 5 major categories: Animals; Natural soundscapes \& water sounds; Human and non-speech sounds; Interior/domestic sounds; Exterior/urban noises. We recommend readers to visit ESC-50's official homepage~\footnote{\url{https://github.com/karolpiczak/ESC-50}}.

\subsection{FSD50K (F50k)}
FSD50K, or the Freesound Dataset 50k, is an open-access repository of human-labeled sound events, comprising 51,197 audio clips sourced from Freesound. These clips span 200 distinct sound classes—drawn from a subset of the AudioSet Ontology—including 144 leaf nodes (specific sound categories) and 56 intermediate nodes (broader taxonomic groupings)—totaling 108.3 hours of multi-labeled audio content. Developed by the Music Technology Group at Universitat Pompeu Fabra, FSD50K serves as a robust resource for advancing research in machine listening, particularly for large-vocabulary sound event classification tasks. We recommend readers to read the paper presenting FSD50K~\footnote{\url{https://arxiv.org/abs/2010.00475}}.

\subsection{Gunshot Triangulation (Gun)}
The Gunshot Triangulation task is one of the tasks in the HEAR benchmark. Its goal is to identify the location of a microphone recording a gunshot using classification. There are a total of 88 clips of data with duration of 1.5 seconds, evaluated by dividing the data into 7 folds. We recommend readers to visit HEAR benchmark official homepage to gain a more detailed understanding of Gunshot Triangulation task~\footnote{\url{https://hearbenchmark.com/hear-tasks.html}}.

\subsection{GTZAN Genre (GZ-Gen)}
The GTZAN Genre task is music genre classification task. The dataset consists of 1000 audio tracks each 30 seconds long. It contains 10 genres, each represented by 100 tracks. The tracks are all 22050Hz Mono 16-bit audio files. The genres contain blues, classical, country, disco, hiphop, jazz, metal, pop, reggae and rock. We recommend readers to visit GTZAN's official homepage~\footnote{\url{http://marsyas.info/index.html}}.

\subsection{GTZAN Music Speech (GZM/S)}
The GTZAN Music Speech task is classification of audio into music or speech task.The dataset was collected for the purposes of music/speech discrimination. The dataset consists of 120 tracks, each 30 seconds long. Each class (music/speech) has 60 examples. The tracks are all 22050Hz Mono 16-bit audio files. We recommend readers to visit GTZAN's official homepage~\footnote{\url{http://marsyas.info/index.html}}.

\subsection{LibriCount (LiCt)}
LibriCount is a synthetic dataset for multiclass speaker count estimation. The dataset contains a simulated cocktail party environment of 0 to 10 speakers, mixed with 0dB SNR from random utterances of different speakers from the LibriSpeech CleanTest dataset. All recordings are of 5s durations, and all speakers are active for the most part of the recording. We recommend readers to visit HEAR benchmark official homepage to gain a more detailed understanding of LibriCount task.

\subsection{Mridingham Stroke and Mridingham Tonic (Mri-S, Mri-T)}
The Mridangam dataset is a collection of 6977 audio examples with duration of 0.81 seconds of individual strokes of the Mridangam in various tonics. The dataset comprises of 10 different strokes played on Mridangams with 6 different tonic values. These two tasks classify stroke and tonic separately. We recommend readers to visit HEAR benchmark official homepage to gain a more detailed understanding of Mridingham Stroke and Mridingham Tonic tasks.

\subsection{Vocal Imitations (VI)}
The VocalImitationSet is a collection of crowd-sourced vocal imitations of a large set of diverse sounds collected from Freesound~\footnote{\url{https://freesound.org/}}, which were curated based on Google's AudioSet ontology~\footnote{\url{https://research.google.com/audioset/}}. This dataset help research communities obtain better understanding of human's vocal imitation and build a machine understand the imitations as humans do. This task is aim to match a vocal imitation to the type of sound imitated, using classification. We recommend readers to visit Vocal Imitations's official homepage~\footnote{\url{https://github.com/interactiveaudiolab/VocalImitationSet}}.

\subsection{VoxLingua107 Top 10 (VL)}
The VoxLingua107 Top 10 task is a new multi class classification task derived from the VoxLingua107 dataset, with the goal of identifying spoken languages in audio files, including 10 spoken languages: Arabic (ar), Danish (da), Estonian (et), Persian (fa), Finnish (fi), French (fr), Armenian (hy), Latvian (lv), Dutch (nl), and Swedish (sv). We recommend readers to read the paper presenting VoxLingua107~\footnote{\url{https://arxiv.org/abs/2011.12998}}.

\section{Details of formula fitting}
We use the \texttt{curve\_fit} function in the open-source \texttt{SciPy} computing library to fit our formulas in the main text. In practical situations, because $Q_\infty$ cannot be lower than 0 or exceed 1, we set the range of parameter $Q_\infty$ to [0,1] when fitting the formulas. For other parameters, we set them to be greater than 0. We set the maximum number of iterations for fitting to 5000. 

In addition, we use the coefficient of determination $R^2$ to measure the degree of fit of the formulas to the data. The calculation method for the coefficient of determination $R^2$ is as follows:
\begin{equation}
\label{eq:R2}
    R^{2}=1- \frac{ \sum \left( y_{i}- \widehat{y}_{i} \right)^{2}}{ \sum \left( y_{i}- \overline{y} \right)^{2}}
\end{equation}
with
\begin{equation}
\overline{y}= \frac{1}{n} \sum y_{i}
\end{equation}
where $y_{i}$ represents the true value (actual observed data), $\widehat{y}_{i}$ represents the predicted value of the fitting function, and $\overline{y}$ represents the mean of the true value. $\sum \left( y_{i}- \widehat{y}_{i} \right)^{2}$ part of Formula~\eqref{eq:R2} represents the Residual Sum of Squares (RSS), which reflects the total amount of model prediction error. $\sum \left( y_{i}- \overline{y} \right)^{2}$ part of Formula~\eqref{eq:R2} represents the Total Sum of Squares (TSS), which reflects the overall variability of the data. $R^2$ is a measure of the proportion of RSS in TSS. The range of $R^2$ values is from 0 to 1, and the higher the $R^2$ value, the better the formulas fit the data.

\bibliographystyle{IEEEtran}
\bibliography{refs}

% Generated by IEEEtran.bst, version: 1.14 (2015/08/26)
\begin{thebibliography}{10}
\providecommand{\url}[1]{#1}
\csname url@samestyle\endcsname
\providecommand{\newblock}{\relax}
\providecommand{\bibinfo}[2]{#2}
\providecommand{\BIBentrySTDinterwordspacing}{\spaceskip=0pt\relax}
\providecommand{\BIBentryALTinterwordstretchfactor}{4}
\providecommand{\BIBentryALTinterwordspacing}{\spaceskip=\fontdimen2\font plus
\BIBentryALTinterwordstretchfactor\fontdimen3\font minus \fontdimen4\font\relax}
\providecommand{\BIBforeignlanguage}[2]{{%
\expandafter\ifx\csname l@#1\endcsname\relax
\typeout{** WARNING: IEEEtran.bst: No hyphenation pattern has been}%
\typeout{** loaded for the language `#1'. Using the pattern for}%
\typeout{** the default language instead.}%
\else
\language=\csname l@#1\endcsname
\fi
#2}}
\providecommand{\BIBdecl}{\relax}
\BIBdecl

\bibitem{kaplan2020scaling}
J.~Kaplan, S.~McCandlish, T.~Henighan, T.~B. Brown, B.~Chess, R.~Child, S.~Gray, A.~Radford, J.~Wu, and D.~Amodei, ``Scaling laws for neural language models,'' \emph{arXiv preprint arXiv:2001.08361}, 2020.

\bibitem{isik2024scaling}
B.~Isik, N.~Ponomareva, H.~Hazimeh, D.~Paparas, S.~Vassilvitskii, and S.~Koyejo, ``Scaling laws for downstream task performance of large language models,'' in \emph{ICLR 2024 Workshop on Mathematical and Empirical Understanding of Foundation Models}, 2024.

\bibitem{hoffmann2022training}
J.~Hoffmann, S.~Borgeaud, A.~Mensch, E.~Buchatskaya, T.~Cai, E.~Rutherford, D.~d.~L. Casas, L.~A. Hendricks, J.~Welbl, A.~Clark \emph{et~al.}, ``Training compute-optimal large language models,'' \emph{arXiv preprint arXiv:2203.15556}, 2022.

\bibitem{muennighoff2023scaling}
N.~Muennighoff, A.~Rush, B.~Barak, T.~Le~Scao, N.~Tazi, A.~Piktus, S.~Pyysalo, T.~Wolf, and C.~A. Raffel, ``Scaling data-constrained language models,'' \emph{Advances in Neural Information Processing Systems}, vol.~36, pp. 50\,358--50\,376, 2023.

\bibitem{aghajanyan2023scaling}
A.~Aghajanyan, L.~Yu, A.~Conneau, W.-N. Hsu, K.~Hambardzumyan, S.~Zhang, S.~Roller, N.~Goyal, O.~Levy, and L.~Zettlemoyer, ``Scaling laws for generative mixed-modal language models,'' in \emph{International Conference on Machine Learning}.\hskip 1em plus 0.5em minus 0.4em\relax PMLR, 2023, pp. 265--279.

\bibitem{isik2025scaling}
B.~Isik, N.~Ponomareva, H.~Hazimeh, D.~Paparas, S.~Vassilvitskii, and S.~Koyejo, ``Scaling laws for downstream task performance in machine translation,'' in \emph{The Thirteenth International Conference on Learning Representations}, 2025.

\bibitem{fan2024scaling}
L.~Fan, K.~Chen, D.~Krishnan, D.~Katabi, P.~Isola, and Y.~Tian, ``Scaling laws of synthetic images for model training... for now,'' in \emph{Proceedings of the IEEE/CVF Conference on Computer Vision and Pattern Recognition}, 2024, pp. 7382--7392.

\bibitem{klug2022scaling}
T.~Klug and R.~Heckel, ``Scaling laws for deep learning based image reconstruction,'' \emph{arXiv preprint arXiv:2209.13435}, 2022.

\bibitem{cherti2023reproducible}
M.~Cherti, R.~Beaumont, R.~Wightman, M.~Wortsman, G.~Ilharco, C.~Gordon, C.~Schuhmann, L.~Schmidt, and J.~Jitsev, ``Reproducible scaling laws for contrastive language-image learning,'' in \emph{Proceedings of the IEEE/CVF conference on computer vision and pattern recognition}, 2023, pp. 2818--2829.

\bibitem{bengio2013representation}
Y.~Bengio, A.~Courville, and P.~Vincent, ``Representation learning: A review and new perspectives,'' \emph{IEEE transactions on pattern analysis and machine intelligence}, vol.~35, no.~8, pp. 1798--1828, 2013.

\bibitem{mohamed2022self}
A.~Mohamed, H.-y. Lee, L.~Borgholt, J.~D. Havtorn, J.~Edin, C.~Igel, K.~Kirchhoff, S.-W. Li, K.~Livescu, L.~Maal{\o}e \emph{et~al.}, ``Self-supervised speech representation learning: A review,'' \emph{IEEE Journal of Selected Topics in Signal Processing}, vol.~16, no.~6, pp. 1179--1210, 2022.

\bibitem{ericsson2022self}
L.~Ericsson, H.~Gouk, C.~C. Loy, and T.~M. Hospedales, ``Self-supervised representation learning: Introduction, advances, and challenges,'' \emph{IEEE Signal Processing Magazine}, vol.~39, no.~3, pp. 42--62, 2022.

\bibitem{li2025speech}
H.~Li, J.~Q. Yip, T.~Fan, and E.~S. Chng, ``Speech enhancement using continuous embeddings of neural audio codec,'' in \emph{ICASSP 2025-2025 IEEE International Conference on Acoustics, Speech and Signal Processing (ICASSP)}.\hskip 1em plus 0.5em minus 0.4em\relax IEEE, 2025, pp. 1--5.

\bibitem{deng2025scaling}
X.~Deng, T.~Wan, K.~Xu, T.~Gao, P.~Qiao, D.~Feng, and Y.~Dou, ``Scaling bioacoustic signal pre-training with million samples via mask-modeling,'' in \emph{ICASSP 2025-2025 IEEE International Conference on Acoustics, Speech and Signal Processing (ICASSP)}.\hskip 1em plus 0.5em minus 0.4em\relax IEEE, 2025, pp. 1--5.

\bibitem{whetten2025towards}
R.~Whetten, L.~Maison, T.~Parcollet, M.~Dinarelli, and Y.~Est{\`e}ve, ``Towards early prediction of self-supervised speech model performance,'' in \emph{Interspeech 2025}, 2025.

\bibitem{he2022masked}
K.~He, X.~Chen, S.~Xie, Y.~Li, P.~Doll{\'a}r, and R.~Girshick, ``Masked autoencoders are scalable vision learners,'' in \emph{Proceedings of the IEEE/CVF conference on computer vision and pattern recognition}, 2022, pp. 16\,000--16\,009.

\bibitem{huang2022masked}
P.-Y. Huang, H.~Xu, J.~Li, A.~Baevski, M.~Auli, W.~Galuba, F.~Metze, and C.~Feichtenhofer, ``Masked autoencoders that listen,'' \emph{Advances in Neural Information Processing Systems}, vol.~35, pp. 28\,708--28\,720, 2022.

\bibitem{cover2006geometrical}
T.~M. Cover, ``Geometrical and statistical properties of systems of linear inequalities with applications in pattern recognition,'' \emph{IEEE transactions on electronic computers}, no.~3, pp. 326--334, 2006.

\bibitem{garrido2023rankme}
Q.~Garrido, R.~Balestriero, L.~Najman, and Y.~Lecun, ``Rankme: Assessing the downstream performance of pretrained self-supervised representations by their rank,'' in \emph{International conference on machine learning}.\hskip 1em plus 0.5em minus 0.4em\relax PMLR, 2023, pp. 10\,929--10\,974.

\bibitem{dinkel2024dasheng}
H.~Dinkel, Z.~Yan, Y.~Wang, J.~Zhang, Y.~Wang, and B.~Wang, ``Scaling up masked audio encoder learning for general audio classification,'' in \emph{Interspeech 2024}, 2024.

\bibitem{gong2022ssast}
Y.~Gong, C.-I. Lai, Y.-A. Chung, and J.~Glass, ``Ssast: Self-supervised audio spectrogram transformer,'' in \emph{Proceedings of the AAAI Conference on Artificial Intelligence}, vol.~36, no.~10, 2022, pp. 10\,699--10\,709.

\bibitem{hsu2021hubert}
W.-N. Hsu, B.~Bolte, Y.-H.~H. Tsai, K.~Lakhotia, R.~Salakhutdinov, and A.~Mohamed, ``Hubert: Self-supervised speech representation learning by masked prediction of hidden units,'' \emph{IEEE/ACM transactions on audio, speech, and language processing}, vol.~29, pp. 3451--3460, 2021.

\bibitem{baevski2020wav2vec}
A.~Baevski, Y.~Zhou, A.~Mohamed, and M.~Auli, ``wav2vec 2.0: A framework for self-supervised learning of speech representations,'' \emph{Advances in neural information processing systems}, vol.~33, pp. 12\,449--12\,460, 2020.

\bibitem{hsu2021robust}
W.-N. Hsu, A.~Sriram, A.~Baevski, T.~Likhomanenko, Q.~Xu, V.~Pratap, J.~Kahn, A.~Lee, R.~Collobert, G.~Synnaeve \emph{et~al.}, ``Robust wav2vec 2.0: Analyzing domain shift in self-supervised pre-training,'' \emph{arXiv preprint arXiv:2104.01027}, 2021.

\bibitem{wang2020fairseq}
C.~Wang, Y.~Tang, X.~Ma, A.~Wu, S.~Popuri, D.~Okhonko, and J.~Pino, ``Fairseq s2t: Fast speech-to-text modeling with fairseq,'' \emph{arXiv preprint arXiv:2010.05171}, 2020.

\bibitem{gulati2020conformer}
A.~Gulati, J.~Qin, C.-C. Chiu, N.~Parmar, Y.~Zhang, J.~Yu, W.~Han, S.~Wang, Z.~Zhang, Y.~Wu \emph{et~al.}, ``Conformer: Convolution-augmented transformer for speech recognition,'' \emph{arXiv preprint arXiv:2005.08100}, 2020.

\bibitem{turian2022hear}
J.~Turian, J.~Shier, H.~R. Khan, B.~Raj, B.~W. Schuller, C.~J. Steinmetz, C.~Malloy, G.~Tzanetakis, G.~Velarde, K.~McNally \emph{et~al.}, ``Hear: Holistic evaluation of audio representations,'' in \emph{NeurIPS 2021 Competitions and Demonstrations Track}.\hskip 1em plus 0.5em minus 0.4em\relax PMLR, 2022, pp. 125--145.

\bibitem{press2007numerical}
W.~H. Press, \emph{Numerical recipes 3rd edition: The art of scientific computing}.\hskip 1em plus 0.5em minus 0.4em\relax Cambridge university press, 2007.

\bibitem{roy2007effective}
O.~Roy and M.~Vetterli, ``The effective rank: A measure of effective dimensionality,'' in \emph{2007 15th European signal processing conference}.\hskip 1em plus 0.5em minus 0.4em\relax IEEE, 2007, pp. 606--610.

\bibitem{shannon1948mathematical}
C.~E. Shannon, ``A mathematical theory of communication,'' \emph{The Bell system technical journal}, vol.~27, no.~3, pp. 379--423, 1948.

\bibitem{chen2020simple}
T.~Chen, S.~Kornblith, M.~Norouzi, and G.~Hinton, ``A simple framework for contrastive learning of visual representations,'' in \emph{International conference on machine learning}.\hskip 1em plus 0.5em minus 0.4em\relax PmLR, 2020, pp. 1597--1607.

\bibitem{bardes2021vicreg}
A.~Bardes, J.~Ponce, and Y.~LeCun, ``Vicreg: Variance-invariance-covariance regularization for self-supervised learning,'' \emph{arXiv preprint arXiv:2105.04906}, 2021.

\bibitem{caron2021emerging}
M.~Caron, H.~Touvron, I.~Misra, H.~J{\'e}gou, J.~Mairal, P.~Bojanowski, and A.~Joulin, ``Emerging properties in self-supervised vision transformers,'' in \emph{Proceedings of the IEEE/CVF international conference on computer vision}, 2021, pp. 9650--9660.

\bibitem{aldeneh2025towards}
Z.~Aldeneh, V.~Thilak, T.~Higuchi, B.-J. Theobald, and T.~Likhomanenko, ``Towards automatic assessment of self-supervised speech models using rank,'' in \emph{ICASSP 2025-2025 IEEE International Conference on Acoustics, Speech and Signal Processing (ICASSP)}.\hskip 1em plus 0.5em minus 0.4em\relax IEEE, 2025, pp. 1--5.

\bibitem{henighan2020scaling}
T.~Henighan, J.~Kaplan, M.~Katz, M.~Chen, C.~Hesse, J.~Jackson, H.~Jun, T.~B. Brown, P.~Dhariwal, S.~Gray \emph{et~al.}, ``Scaling laws for autoregressive generative modeling,'' \emph{arXiv preprint arXiv:2010.14701}, 2020.

\bibitem{droppo21_interspeech}
J.~Droppo and O.~Elibol, ``Scaling laws for acoustic models,'' in \emph{Interspeech 2021}, 2021, pp. 2576--2580.

\bibitem{touvron2023llama}
H.~Touvron, T.~Lavril, G.~Izacard, X.~Martinet, M.-A. Lachaux, T.~Lacroix, B.~Rozi{\`e}re, N.~Goyal, E.~Hambro, F.~Azhar \emph{et~al.}, ``Llama: Open and efficient foundation language models,'' \emph{arXiv preprint arXiv:2302.13971}, 2023.

\bibitem{zhu2024multichannel}
Q.~Zhu, J.~Zhang, Y.~Gu, Y.~Hu, and L.~Dai, ``Multichannel av-wav2vec2: A framework for learning multichannel multi-modal speech representation,'' in \emph{Proceedings of the AAAI Conference on Artificial Intelligence}, vol.~38, no.~17, 2024, pp. 19\,768--19\,776.

\bibitem{zhu2025muq}
H.~Zhu, Y.~Zhou, H.~Chen, J.~Yu, Z.~Ma, R.~Gu, Y.~Luo, W.~Tan, and X.~Chen, ``Muq: Self-supervised music representation learning with mel residual vector quantization,'' \emph{arXiv preprint arXiv:2501.01108}, 2025.

\bibitem{shams2024ssamba}
S.~Shams, S.~S. Dindar, X.~Jiang, and N.~Mesgarani, ``Ssamba: Self-supervised audio representation learning with mamba state space model,'' in \emph{2024 IEEE Spoken Language Technology Workshop (SLT)}.\hskip 1em plus 0.5em minus 0.4em\relax IEEE, 2024, pp. 1053--1059.

\bibitem{zaman2023survey}
K.~Zaman, M.~Sah, C.~Direkoglu, and M.~Unoki, ``A survey of audio classification using deep learning,'' \emph{IEEE access}, vol.~11, pp. 106\,620--106\,649, 2023.

\bibitem{zhou2025audio}
J.~Zhou, X.~Shen, J.~Wang, J.~Zhang, W.~Sun, J.~Zhang, S.~Birchfield, D.~Guo, L.~Kong, M.~Wang \emph{et~al.}, ``Audio-visual segmentation with semantics,'' \emph{International Journal of Computer Vision}, vol. 133, no.~4, pp. 1644--1664, 2025.

\bibitem{chen2019audio}
Y.-C. Chen, S.-F. Huang, H.-y. Lee, Y.-H. Wang, and C.-H. Shen, ``Audio word2vec: Sequence-to-sequence autoencoding for unsupervised learning of audio segmentation and representation,'' \emph{IEEE/ACM Transactions on Audio, Speech, and Language Processing}, vol.~27, no.~9, pp. 1481--1493, 2019.

\bibitem{yuan2025sound}
Y.~Yuan, D.~Jia, X.~Zhuang, Y.~Chen, Z.~Chen, Y.~Wang, Y.~Wang, X.~Liu, X.~Kang, M.~D. Plumbley \emph{et~al.}, ``Sound-vecaps: Improving audio generation with visually enhanced captions,'' in \emph{ICASSP 2025-2025 IEEE International Conference on Acoustics, Speech and Signal Processing (ICASSP)}.\hskip 1em plus 0.5em minus 0.4em\relax IEEE, 2025, pp. 1--5.

\bibitem{yang2023uniaudio}
D.~Yang, J.~Tian, X.~Tan, R.~Huang, S.~Liu, X.~Chang, J.~Shi, S.~Zhao, J.~Bian, X.~Wu \emph{et~al.}, ``Uniaudio: An audio foundation model toward universal audio generation,'' \emph{arXiv preprint arXiv:2310.00704}, 2023.

\bibitem{gemmeke2017audio}
J.~F. Gemmeke, D.~P. Ellis, D.~Freedman, A.~Jansen, W.~Lawrence, R.~C. Moore, M.~Plakal, and M.~Ritter, ``Audio set: An ontology and human-labeled dataset for audio events,'' in \emph{2017 IEEE international conference on acoustics, speech and signal processing (ICASSP)}.\hskip 1em plus 0.5em minus 0.4em\relax IEEE, 2017, pp. 776--780.

\bibitem{lee2021acav100m}
S.~Lee, J.~Chung, Y.~Yu, G.~Kim, T.~Breuel, G.~Chechik, and Y.~Song, ``Acav100m: Automatic curation of large-scale datasets for audio-visual video representation learning,'' in \emph{Proceedings of the IEEE/CVF International Conference on Computer Vision}, 2021, pp. 10\,274--10\,284.

\bibitem{anton2023audio}
J.~Anton, H.~Coppock, P.~Shukla, and B.~W. Schuller, ``Audio barlow twins: Self-supervised audio representation learning,'' in \emph{ICASSP 2023-2023 IEEE International Conference on Acoustics, Speech and Signal Processing (ICASSP)}.\hskip 1em plus 0.5em minus 0.4em\relax IEEE, 2023, pp. 1--5.

\end{thebibliography}

% \newpage

\vspace{-25pt}
\begin{IEEEbiographynophoto}{Xuyao Deng} received the bachelor's degree from Shandong University, Shandong, China, in 2023. He is currently a doctoral student at the College of Computer Science and Technology, National University of Defense Technology, Changsha, China. His research interests include signal processing, machine learning, and intelligent software systems.
\vspace{-25pt}
\end{IEEEbiographynophoto}

\begin{IEEEbiographynophoto}{Kele Xu} (Senior Member, IEEE) received the doctorate degree from Paris VI University, Paris, France, in 2017. He is currently an Associate Professor with the College of Computer Science and Technology, National University of Defense Technology, Changsha, China. His research interests include audio signal processing, machine learning, and intelligent software systems.
\vspace{-25pt}
\end{IEEEbiographynophoto}

\begin{IEEEbiographynophoto}{Yong Dou} Professor and Ph.D. supervisor with the National Key Laboratory of Parallel and Distributed Computing, National University of Defense Technology. His research interests cover high performance computing, intelligent computing, machine learning, and deep learning.
\end{IEEEbiographynophoto}

\vfill

\end{document}